# Stabilizing skyrmions in stepped magnetic devices for multistate memory

W. Al Saidi,[1] R. Sbiaa,* S. Al Risi,[1] F. Al Shanfari[1], N. Tiercelin[2] and Y. Dusch[2]

[1]Sultan Qaboos University, Department of Physics, P.O. Box 36, PC 123, Muscat, Oman

[2]University of Lille, CNRS, Centrale Lille, Université Polytechnique Hauts-de-France, UMR 8520 - IEMN, 59000 Lille, France

* Corresponding author

The dynamics and stability of magnetic skyrmions within a nano-track with multiple confinements are investigated. Firstly, the motion of a single skyrmion under spin transfer torque (STT) is studied. By accurately adjusting the current pulse magnitude and width, the study reveals the possibility to pin and stabilize the skyrmion in each confinement. Due to the Hall angle, the depining of the skyrmion from the top confinement requires two pulses with adjustable time delay while a single pulse is enough to depin it for the case of bottom confinement. In the case of two skyrmions, once one is pinned in one confinement, the second one stabilizes in the nearest available empty state and no more than one skyrmion could be seen in single confinement. Finally and for further confirmation of this behavior, the motion of a large number of skyrmions is investigated under the same conditions. The results show that a multistate device could be obtained with still the existence of only one skyrmion per state. The skyrmions could be displaced along the nano-track until their annihilation at the end of the device.

## 1. Introduction

Since the first experimental observations of skyrmions in MnSi [1,2] and thin ferromagnetics at room temperature [3,4], intensive studies have been focused on understanding their behavior. The nanoscale size and stability of these quasi-particle magnetic structures make them attractive for their implementation in different applications such as memory [5-9], logic [10] and neuromorphic computing [11-15]. Recently, it has been demonstrated that skyrmions can be created, displaced and annihilated by applying a magnetic field [16,17], spin transfer torque (STT) [18-20], spin-orbit torque [21], current and ultrafast heating pulses [22-24] or even surface acoustic waves [25]. It is worth noting that skyrmions can be displaced with current density with values that could be better than $10^{11}$ A/m$^2$ [26] which is much lower than what is required for moving a magnetic domain wall case [27,28]. It has been reported that controlling the materials intrinsic properties provides an easy way to tune the skyrmions size and their velocity [29-32] which are among the key characteristics of a functional device; i.e. small size leads to higher memory capacity while high velocity reflects the speed of changing and writing the data. Although the motion of skyrmions within nano-tracks has been widely investigated, their accurate positioning remains a challenge and has not been deeply investigated. In our previous work, we showed that creating stepped confinement could help to pin and stabilize a skyrmion [32]. In this paper, we demonstrate the possibility to control the position of a large number of skyrmions in a multi-stepped nanowire. More interestingly, the study revealed that in the proposed design only a single skyrmion could be stabilized in each confinement. The flow of a chain of skyrmions could be properly controlled along the nanowire. The effect of the physical barrier and the magnetic force between the skyrmions play a major role in the observed behavior.

**2. Theoretical Model**

The theoretical model of multistate device generation is performed by micromagnetic simulation using Mumax3 software [33] that is based on solving the Landau–Lifshitz–Gilbert (LLG) Equation without temperature effect (T = 0 K). For the case of current flowing in the plane of the film (CIP), the LLG equation can be written as:

$$\frac{d\vec{m}}{dt} = -\gamma \, \vec{m} \times \vec{H}_{eff} + \alpha \, (\vec{m} \times d\vec{m}/dt) + \vec{\Gamma}_{SST} \qquad (1)$$

The first and the second term of the right side of Equation 1 are the precession and damping terms respectively. The effective magnetic field is expressed as:

$$\vec{H}_{eff} = \frac{2A}{\mu_o M_S} \nabla^2 \vec{m} + \frac{2K_u}{\mu_o M_S} m_z \hat{z} + \vec{H}_D + \vec{H}_{DMI} \qquad (2)$$

The terms on the right side of Equation 2 are the exchange field, the crystalline magnetic anisotropy field, the demagnetization field and the DMI field, respectively. The last term of the LLG Equation 1 is associated with the spin transfer torque (STT) effect in the case of a current flowing in the plane of the device and is expressed by Zhang-Li form [34]

$$\vec{\Gamma}_{SST} = -(\vec{v}_S.\vec{\nabla})\vec{m} + \beta\,\vec{m} \times (\vec{v}_S.\vec{\nabla})\vec{m} \qquad (3)$$

where $v_S$ is the effective spin-current drift velocity and β is the non-adiabatic spin torque parameter. The main objective of this study is to investigate a novel design for controlling the motion of the skyrmion in a memory device. We consider a 2 nm thick nano-track with a length of 780 nm and a width of 80 nm. Stepped regions with a size of 30 × 70 nm² were added and removed from some parts of the track to localize the skyrmion position and to improve the linearity and repeatability of storage devices. The material intrinsic parameters used for this study are saturation magnetization $M_S$ = 500 kA/m, exchange stiffness $A_{ex}$ = 15 pJ/m, Dzyaloshinskii Moriya interaction constant $D$ = 3.3 mJ/m², Gilbert-damping coefficient α = 0.1 and the perpendicular magnetic anisotropy (PMA) $K_u = 0.8 \times 10^6$ J/m³. The shape anisotropy originated from the dipole-dipole interactions and causes a shift in the constant of uniaxial anisotropy in ultrathin film cases where the non-local effects are negligible [35,36]. The sample is discretized into 512 × 128 × 1 cells, corresponding to a unit cell size of 2 × 2 × 3 nm³. The values of $M_S$, $K_u$ $A_{ex}$ and α are characteristics of (Co/Ni), (Co/Pt) and (Co/Pd) multilayers [36-40]. They can be tuned in a wide range by adjusting the respective thicknesses and the number of repeats. The value of $D$ in typical asymmetric structures where one Pt at the interface is substituted by Ir, W, AlOx for instance [41-43]. The magnetic parameters used in this paper are similar to our previous work [32,44]

Firstly, a Néel-type skyrmion with topological charge Q= -1 is created near the left edge of the nano-track and then an electric current with a spin polarization efficiency *P* fixed to 0.4 is injected into the nano-track in the CIP geometry to generate a motion of the skyrmion. The non-adiabatic torque coefficient $\beta$ is fixed to 0.2 and the current flows toward the left so the electron would flow in the opposite direction.

## 3. Results and discussion

When the skyrmion is present in a wire it represents bit "1" while its absence at a particular position represents bit "0". To avoid information loss, the position of the skyrmion has to be fixed at a predefined position and the annihilation has to be avoided; hence in the present work, a constructed wire is used to create pinning regions for the skyrmion to define bit positions. Each region represents one bit, which corresponds to "0" if the region does not contain a skyrmion-bubble and "1" if the region contains one skyrmion-bubble.

### 3.1. Dynamics of a single skyrmion in constricted nanowire

In the first part of this study, the pinning and reliable depinning of the skyrmion from the pinning centers under current pulses to have a multistate memory device has been investigated. It can be seen from Figure 1 that it is possible to achieve 8 states by pinning the skyrmion in the confinement regions. The calculations were carried out by creating a skyrmion on the left edge as the initial state and a sequence of current pulses were applied. Both the pulse width and the time delay between the pulses were fixed at 2 ns while the current density was varied. Figure 1(a) shows the position of the skyrmion versus time. At the right side of the graph, the magnitude of the current density is indicated. The first state was obtained for a pulse with magnitude $J = 6.0 \times 10^{11}$ A/m² indicated by number 1 in Figure 1(b). It can be seen that the skyrmion moves with a constant velocity during the period of 2 ns when the current is applied as indicated by an almost straight line until it reaches a position of about 120 nm then bounces back and stabilizes. Skyrmion shows noticeable inertia-driven drift after the current pulse is removed rather than stopping immediately. The color scale in Figure 1(b) represents the out-of-plane component $m_z$ of the magnetization. The displacement of the skyrmion to state 2 (first top confined region) starting from the initial position can be realized by another single pulse with the same duration and $J = 3.0 \times 10^{11}$ A/m². The slight attenuated oscillations after reaching the maximum position indicate that the skyrmion is swirling within the first confinement before stabilizing as shown in the corresponding snapshot. Moving from state 2 to state 3 requires two pulses, the first one to excite the skyrmion while the second one to move it to the third state. Here we used two pulses with the same magnitude of $5.5 \times 10^{11}$ A/m$^2$. It is important to mention that the need for two pulses to move the skyrmion from the top confinements is due to the spin Hall effect. It is known that the spin Hall effect which is a deviation of the skyrmion path to the edge represents a challenge for skyrmion-based structures [45-48]. To overcome this issue, antiferromagnetically coupled structure [49-52] and spin-orbit torque combined with STT [53] have been proposed to reduce skyrmion deviation to the edge. The other states could be obtained by altering the amplitude and number of pulses as indicated in Figure 1. Interestingly, in order to pin the skyrmion in the upper confinement area (states 2, 4, 6 and 8), a single pulse must be added to the pulses, whereas two pulses are required to move it to the lower confinement area (states 3, 5 and 7). Hence to reach state 8, it was necessary to apply a total of 11 pulses, see Figure 1(c) for the number of pulses required to move the skyrmion to the 11$^{th}$ state. Moreover, the pulse amplitude for the upper confinement was found to be fixed at $3.0 \times 10^{11}$ A/m$^2$ while in the lower confinement, the amplitude is changing depending on the history of the skyrmion. From Figure 1 (a) insert, the oscillation in the $x$ position with time was observed each time the skyrmion is driven from upper to lower states. For more details on the skyrmion dynamics during its motion from one state to the other, the position of the skyrmion versus time has been analyzed.

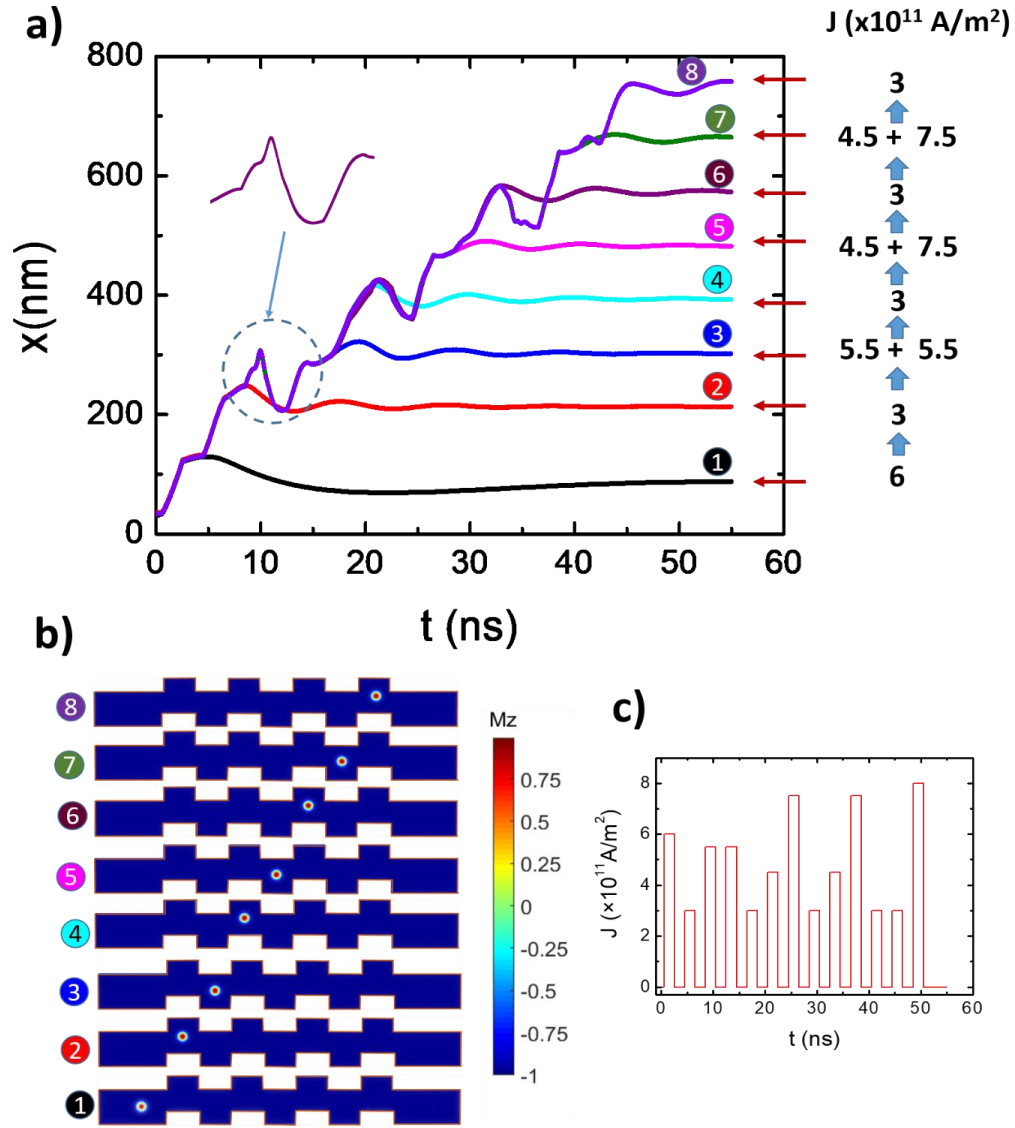

**Figure 1.** (a) Time dependence of the skyrmion position at different current pulse values. After each pulse, the time dependence of the position $x$ is recorded. The following pulse is applied by considering the skyrmion at the initial state. The current density magnitude is shown on the right side. To reach states 3, 5 and 7, two pulses were necessary. The inset is an enlarged $x$ position of the skyrmion showing that the skyrmion stabilized at the top confined region and swirled in that region before moving to the next bottom confinement. (b) snapshots of magnetic states indicated in the left Figure. The length of the wire is $L = 780$ nm, the width $w = 80$ nm and the thickness is 2 nm. The color scale shows the out-of-plane component of magnetization. (c) the electrical current density as a function of time.

The effect of nanowire geometry on current-induced skyrmion dynamics was investigated. Figure 2(a) is a plot of the skyrmion trajectory from state 1 to state 2, it is clear that the skyrmion is not moving in a straight line (blue line). The Magnus force acts in the direction normal to the drift velocity of the skyrmion and can therefore cause spiraling trajectories. The dashed lines indicate the border of the magnetic region. Figure 2(b-e) shows the positions of the skyrmion within the nanowire at different times indicated by numbers 1 to 4 in Figure 2(a). The sign reversal leads to the opposite direction of the topological Magnus force, We found that the skyrmion rotation clockwise and counterclockwise depends on the sign of the skyrmion charge. For the case of a Néel-type skyrmion with -1 charge, the motion is counterclockwise while for +1 the direction of the skyrmion is clockwise. The radius of the path changes depending on the value of the Landau-Lifshitz damping constant that affects the skyrmion Hall effect and the speed [54]. Here only one pulse is applied to move the skyrmion from state 1 to another stable state at the top confined region. Along its path, it was observed that the skyrmion size can be reduced due to the higher compression of skyrmion by the barriers. It was observed that the size of the skyrmion was changing and reduced by about 1 nm at position 2 before expanding by about 2nm when stabilized due to the dipolar repulsive force from the edge [55]. The repulsive force from the edge causes the size of the skyrmion to decrease while passing through the states, if this force is too large in the case of a larger skyrmion or narrower channel for the skyrmion or larger current density, the skyrmion will annihilate. Both the intrinsic parameter and the geometry of the device need to be taken into account while designing the wire, more details can be found in Ref. 32.

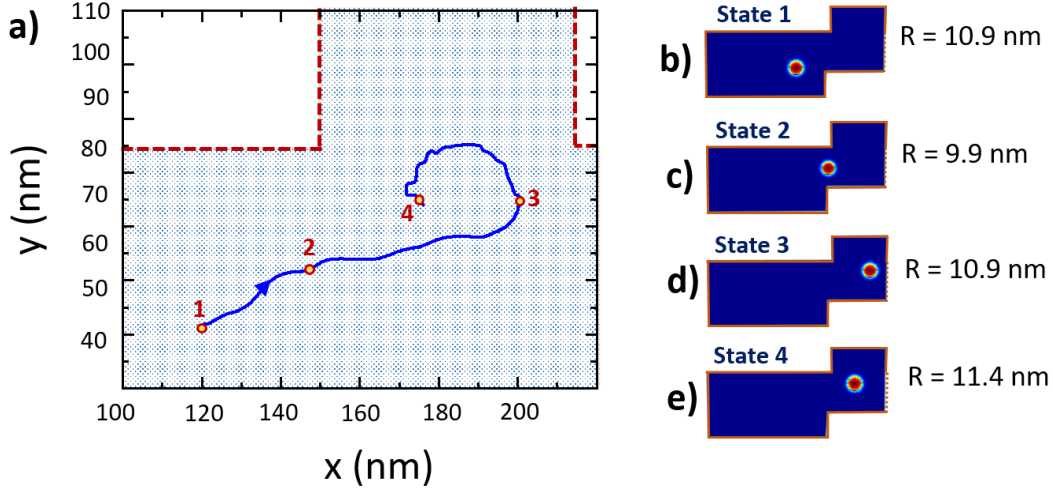

**Figure 2.** The skyrmion trajectory from state 1 to state 2 shows counter-clockwise spiral motion (a) the position of an isolated skyrmion at different times under a pulsed current of $3.0 \times 10^{11}$ A/m² for 2 ns duration (b-e) snapshots showing a part of the device at different positions corresponding to the positions shown in (a).

To move the skyrmion from state 2 to state 3 shown in Figure 1, two pulses were used. The first one with $J = 5.5 \times 10^{11}$ A/m² and 2 ns pulse width for exciting the skyrmion and the second one with the same J and pulse width for displacing it to another state. The dashed area shown inFigure 3 represents a part of the magnetic device where the skyrmion initially at position 1 has been under the two pulses. From Figures 2 and 3, it has been observed that the skyrmion is swirling in 2 different scenarios: i) Firstly, with the displacement from the first state to the second state the skyrmion exhibits a counter-clock motion and then reaches its stable state due to the confinement region (top area). ii) Secondly, moving the skyrmion from the second state to the third one required two pulses. The first one to excite it (will let the skyrmion swirl in the top region) and the second pulse to move the skyrmion to the third confinement region. The skyrmion swirled again counterclockwise before reaching the stable state. The time delay $t_d$ between the two pulses is an important parameter for skyrmion dynamics. The amplitude and width of the two pulses were fixed to $5.5 \times 10^{11}$ A/m$^2$ and 2 ns, respectively while $t_d$ was varied. For $t_d$ larger than 2.8 ns, even with two pulses, the skyrmion remains pinned at state 2 as can be seen in Figure 4(a). Reducing $t_d$ below 2.7 ns and above 1 ns, the skyrmion can be displaced to state 3 shown in Figure 4(b). Finally and for a very short time delay ($t_d < 1$ ns), the skyrmion will be released from its initial state but be annihilated during its motion [Figure 4(c)]. In the first case, $t_d$ is so large that the skyrmion is no more excited by the first pulse and becomes stable before the second pulse is applied. In the last case, it is like having a single pulse with a larger amplitude.

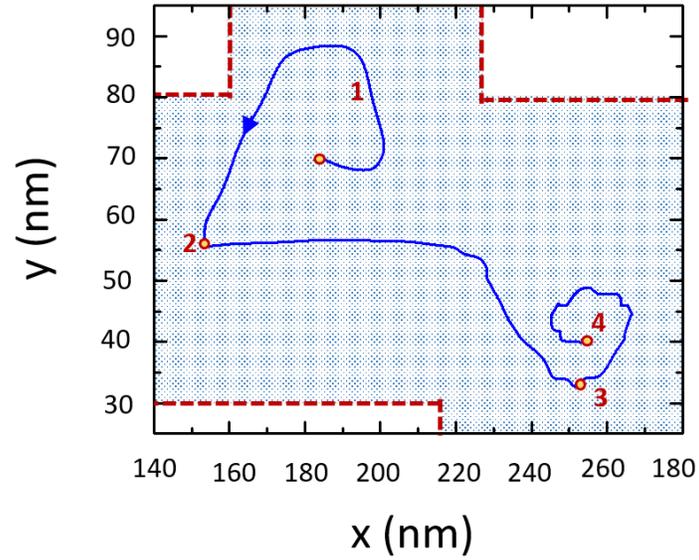

**Figure 3.** The skyrmion trajectory from state 2 to state 3 shows counterclockwise spiral motion. The numbers indicate the positions of the skyrmion at different times under two current pulses of $5.5 \times 10^{11}$ A/m$^2$ for 2 ns duration each and 2 ns time delay between them.

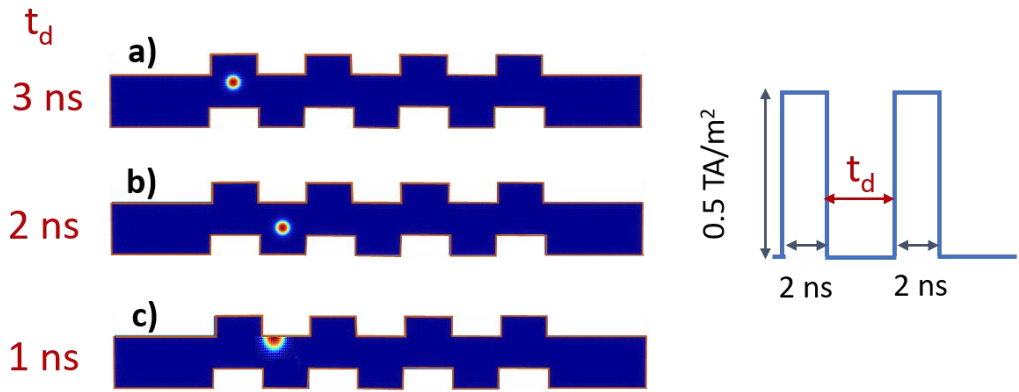

**Figure 4.** (a) The skyrmion remains pinned when the time delay is larger than 2.8 ns, (b) the skyrmion moves to the third state when the time delay is between 1 and 2.7 ns and (c) annihilation of skyrmion when the time delay is less than 1 ns for two fixed pulse of $5.5 \times 10^{11}$ A/m$^2$ magnitude and 2 ns duration.

As discussed earlier, it is possible to stabilize the skyrmion at each state defined in advance by the confined region. For a functional memory application, each state needs to be accessed directly. Figure 5, shows that by changing only the pulse width $\tau$, it is possible to access any state within the nanowire without being pinned in specific confinement. Similar results can be obtained by changing only the pulse magnitude or both of them. For a fixed same current density $J$ and by varying $\tau$, the skyrmion can be displaced at any desired position (state) within the nanowire. For instance at $J = 6.0 \times 10^{11}$ A/m$^2$, the skyrmion can be stabilized at state 3 for $\tau = 4$ ns. Other states such as 5, 8 and 9 can be reached for $\tau$ of 8, 12 and 14 ns, respectively. For memory applications, the optimization of the correct pulse width is essential to avoid written error since for a storage application the position of the skyrmion must be well defined. The current pulse must have sufficient width and strength to move the skyrmion to the desired state. When the current pulse width and strength increase, the probability that the skyrmion moves too far over the next state increases and when it is too low, the skyrmion can not move over the first pinning site. The skyrmion can move back and forth by changing the direction of the current.

**3.2. Two skyrmions dynamics in constricted nanowire**

In the second part of this study, we considered the case of two skyrmions in a chain, initially located at the left side of the nanowire and the focus was on the possibilities of achieving higher numbers of states. We started with two skyrmions namely A (trailing) and B (leading) and we investigated their motion under a pulsed current similar to a single skyrmion case previously discussed. Initially, skyrmion B is leading A by around 40 nm. Under the same pulse with $J = 4.0 \times 10^{11}$ A/m$^2$ and a width of 2 ns, only the leading skyrmion could be displaced to state 2 (top confined region) while the trailing one still remains at state 1 by taking over the position of its neighbor.

### 3.3. Chain of skyrmions in a constricted nanowires

In the third part of this study, we considered a large number of skyrmions moving from the left pad. Very dense encoding information and different states can be achieved using a chain of skyrmions, thus in Figure 6, two large pads have been added at the edges of the nanowire with the size of $150 \times 150 \times 2$ nm$^3$ to enable the creation of a larger number of skyrmions to study a sequence of skyrmions. We initially started with 9 Néel-type skyrmions at the left pad. Eleven pulses with 2ns pulse width were applied to move them along the nanowire while the magnitude of the current density was varied. The resulting phase diagram, shown in Figure 6(a), depicts the number of skyrmions remaining at different regions of the stepped nanowire and also the number of the annihilated skyrmions for different applied current density amplitudes. Figure 6(b-o) shows snapshots of skyrmions within the nanowires at different positions by increasing the current density magnitude with a step of $0.5 \times 10^{11}$ A/m$^2$ after the pulses are applied. For $J < 3.5 \times 10^{11}$ A/m$^2$, it can be seen that after applying current pulses, some skyrmions are moving whereas others remain pinned due to larger pinning potential. In higher driving current, they depin from the confinement since the Lorentz force is high enough [56]. Eleven pulses are not sufficient to lead the skyrmions to the right pad while at $J > 3.5 \times 10^{11}$ A/m$^2$ the skyrmions were displaced at different stable positions and could reach the end of the device. For $J \leq 4.5 \times 10^{11}$ A/m$^2$ the skyrmions move through the pinning site and get blocked without being annihilated. At larger current density, starting from $J = 4.6 \times 10^{11}$ A/m$^2$, the skyrmions are pushed to the edge of the device resulting in their annihilation. This feature is important for the application of the skyrmion-based magnetic racetrack because a smaller current density of depinning means less energy consumption and less Joule heating induced by the current. However, for $J = 3.5 \times 10^{11}$ A/m$^2$, one skyrmion was annihilated since the pulse is given when the skyrmion was near the edge. It is worthy to note that for $J \geq 4.6 \times 10^{11}$ A/m$^2$, some skyrmions are annihilated in the left pad, between the states and even after passing to the right pad. Under these conditions, it is interesting to note that there are some blocked skyrmions at the left pad, no annihilation between skyrmions was observed. More interestingly, only one skyrmion per state is found for this investigated design. By looking very carefully at the skyrmions dynamics, once a skyrmion is reaching one confined region, the trailing one is slowing down and stabilizes at the previous state. A change in the size of the confinement was made and it has been found that increasing the size of the confinement and the gap between them can lead more skyrmions to get pinned in one state, hence designing the device plays an important role to get one skyrmion per state. It has been shown by Bhattacharya *et al.* that it is possible to manipulate fixed skyrmions using voltage-controlled magnetic anisotropy, the isolated skyrmions are created or annihilated by applying voltage pulses that increase or decrease the perpendicular magnetic anisotropy of the system and thus this method could be used to create more skyrmions from the left pad which can be connected to an electrode [57]. Furthermore, another method to write the data is by using the spin Hall currents [58].

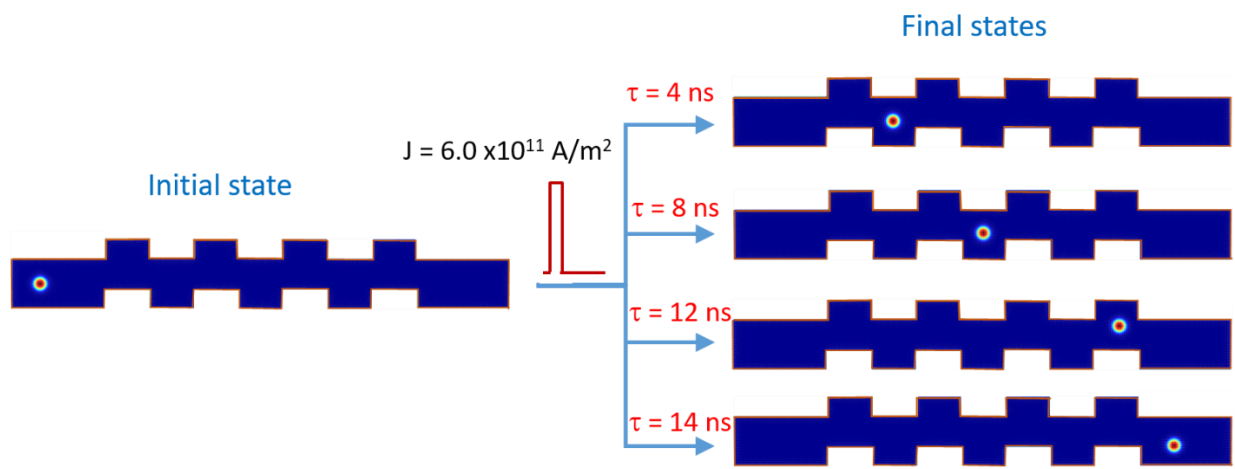

**Figure 5.** Stabilizing a single skyrmion at different states by applying one pulse of $6.0 \times 10^{11}$ A/m$^2$ and varying the pulse width.

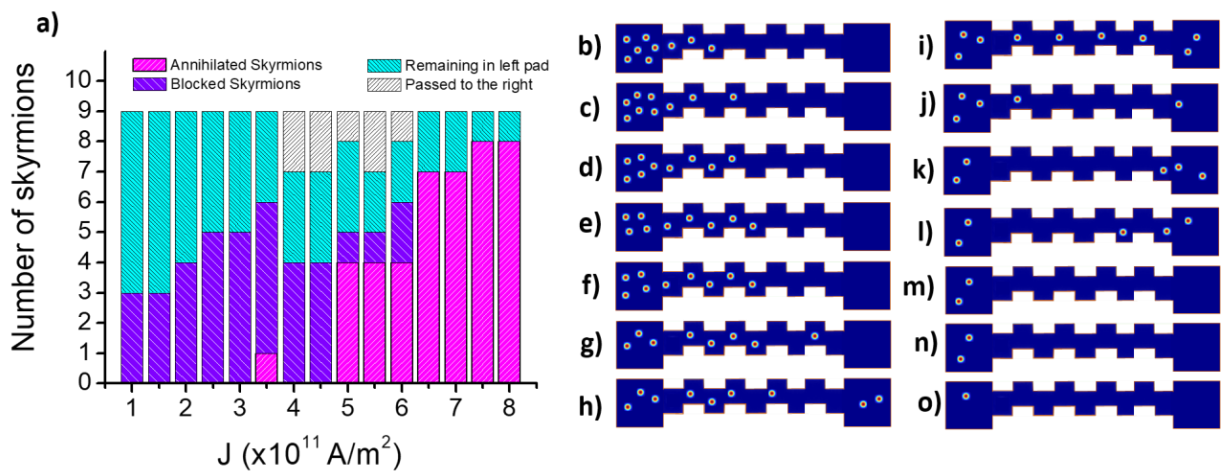

**Figure 6.** The proposed device consists of two magnetic pads and nine injected skyrmions. The net drift of the skyrmion due to eleven current pulses with 2 ns pulse width flowing along the nanowire leads to pinning and annihilation. (a) phase diagram (b-i) different snapshots for different current densities starting from $1.0 \times 10^{11}$ to $8.0 \times 10^{11}$ A/m$^2$ with a step of $0.5 \times 10^{11}$ A/m$^2$.

From Figure 6, it is striking to notice the following: (i) only one skyrmion can occupy a confined region and (ii) by enlarging the right pad and measuring the electrical signal it is possible to count the number of skyrmions reaching the end of the device. The Hall resistivity [59] and measuring the tunnel magnetoresistance [60] can be used to accurately measure the number of passing skyrmions.

## 4. Summary

Towards a multistate skyrmionic device, a stepped nano-track with bottom and top confinements has been proposed. For a common material with perpendicular magnetic anisotropy, the adjustment of current magnitude and its duration enables the displacement of the skyrmion at any position. Furthermore, the study of two skyrmions shows the possibility to move them accurately with only one skyrmion in each confinement. Due to Hall angle, depinning a skyrmion from bottom confinement needs only a single pulse current with an optimal magnitude and duration while two pulses are necessary for depinning from top confinement. If the time delay $t_d$ between the two pulses is too short, the skyrmion collapses and if $t_d$ is too large the skyrmion remains pinned. The dynamic of a large number of skyrmions created at one side shows the possibility to stabilize/pin a single skyrmion in each confinement. These results open the way for large-capacity memory devices and neuromorphic computing.

**Acknowledgments**

The authors would like to acknowledge the support from HMTF Strategic Research of Oman (grant no. SR/SCI/PHYS/20/01).